\documentclass{webofc}

\usepackage[varg]{txfonts}  
\usepackage{hyperref}
\usepackage{url}
\usepackage{lineno}

\usepackage[marginal]{footmisc}
\usepackage{subfigure}
\usepackage{subcaption}

\hypersetup{colorlinks=true,citecolor=blue,urlcolor=blue,linkcolor=blue}

\begin{document}

\title{Measurements of global and local spin polarization of $\Lambda$ and $\bar{\Lambda}$ in Au+Au collisions from the RHIC Beam Energy Scan}

\author{\firstname{Qiang} \lastname{Hu}\inst{1,2} \fnsep\thanks{\email{qianghu@impcas.ac.cn}} for the STAR Collaboration}

\institute{Institute of Modern Physics, Chinese Academy of Sciences, Lanzhou 730000, China 
\and
        School of Nuclear Science and Technology, University of Chinese Academy of Sciences, Beijing
100049, China}

\abstract{We report the measurements of $\Lambda$ hyperons' global and local spin polarization from second phase of the RHIC Beam Energy Scan (BES-II) in Au+Au collisions at $\sqrt{s_{NN}}$= 7.7--27 GeV. Global polarization measurements of $\bar{\Lambda}$ and $\Lambda$ show no significant differences, offering insights into the late-stage evolution of the magnetic field. 
The new measurements of the local polarization of $\Lambda$ perpendicular to the reaction plane ($\langle P_{2,y} \rangle$) shows a monotonic increase with decreasing collision energy, while the component along the beam direction ($\langle P_{2,z} \rangle$) for both $\Lambda$ and $\bar{\Lambda}$ is small in magnitude with no strong energy dependence. The net local polarization observable, $\langle P_{2,y}^{net} \rangle =\langle P_{2,y}(\Lambda) \rangle - \langle P_{2,y}(\bar{\Lambda})\rangle$ and $\langle P_{2,z}^{net} \rangle = \langle P_{2,z}(\Lambda) \rangle - \langle P_{2,z}(\bar{\Lambda} \rangle)$ designed to probe baryonic spin Hall effect, is consistent with zero with large uncertainty.}

\maketitle

\section{Introduction}
\label{intro}
Relativistic heavy-ion collisions provide an excellent opportunity to examine the property of the quark-gluon plasma (QGP) in the laboratory. 
In non-central heavy-ion collisions, the system carries large initial orbital angular momentum. 
Subsequently, the quarks and the final state hadrons with  
non-zero spin could be polarized along global angular momentum due to spin-orbit coupling~\cite{liang2005, voloshin_arxiv}, which is known as global polarization.
Due to parity violation in the decay of a $\Lambda$ hyperon, the daughter proton tends to emit along the spin direction of its parent, making $\Lambda$s excellent candidates for measuring polarization in heavy-ion collisions.

The global polarization is determined by $\left<P_{y}\right>=\frac{8}{\pi\alpha_{\Lambda}}\frac{1}{R_{EP}^{(1)}}\left<\sin(\Psi_{1}-\phi_{p}^{*})\right>$, where $\alpha_{\Lambda}$ is the decay parameter, $\Psi_{1}$ is the first-order of event plane angle and $R_{EP}^{(1)}$ its resolution, and $\phi_{p}^{*}$ is the azimuthal angle of baryon (proton) in $\Lambda$'s rest frame.
The STAR Collaboration carried out the polarization measurement of $\Lambda$ hyperons in Au+Au collisions at $\sqrt{s_{NN}} = 62.4$ and 200 GeV, where the signal was consistent with zero within statistical uncertainties \cite{star_prc024915}. However, the data from first phase of the RHIC Beam Energy Scan (BES-I) provided the first evidence of non-zero global polarization for $\Lambda$ hyperons \cite{star-nature2017, chen2024}. It supports the presence of vortices inside QGP and is considered as one of the important milestones in heavy-ion collisions. The BES-I results also show hints of a difference in global polarization between $\bar{\Lambda}$ and $\Lambda$ which is expected from the effect of the late-stage magnetic field sustained by the QGP. Subsequently, global polarization has been measured by different collaborations in different systems from a few GeV to TeV.

The distribution of polarization as a function of azimuthal angle is referred to as local polarization. Local polarization can be measured with the polarization axis perpendicular to the reaction plane (along the out-of-plane, $\left<P_{2,y}\right>$) and along the reaction plane (along in-plane or beam-axis, $\left<P_{2,z}\right>$), these components are written as $\left<P_{2,y}\right>=\frac{8}{\pi\alpha_{\Lambda}}\frac{1}{R_{EP}^{(1)}}\left<\sin(\Psi_{1}-\phi_{p}^{*})\cos(2\phi_{\Lambda}-2\Psi_{2})\right>$ and $\left<P_{2,z}\right>=\frac{\left<\cos\theta_{p}^{*}\sin(2\phi_{\Lambda}-2\Psi_{2})\right>}{\alpha_{\Lambda}\left<(\cos\theta_{p}^{*})^{2}\right>}$,  where $\Psi_{2}$ is the second-order of event plane angle,
$\theta_p^{*}$ is the polar angle of daughter proton in the $\Lambda$ rest frame relative to the beam direction, $\phi_{\Lambda}$ is the azimuthal angle of $\Lambda$.

The recently predicted baryonic spin Hall effect (SHE)~\cite{liu2021,fu2021,fu2021_she} in heavy-ion collisions describes the splitting of particles with spin up and spin down, driven by the gradient of the baryon chemical potential, similar to how an electric field induces the traditional spin Hall effect in condensed matter physics. The net local polarization, defined as the polarization difference between $\Lambda$ and $\bar{\Lambda}$, denoted by $\langle P_{2,y}^{net} \rangle = \langle P_{2,y}(\Lambda)\rangle - \langle P_{2,y}(\bar{\Lambda})\rangle$ and $\langle P_{2,z}^{net} \rangle = \langle P_{2,z}(\Lambda)\rangle - \langle P_{2,z}(\bar{\Lambda})\rangle$, is predicted to be a sensitive probe for SHE.

\vspace*{-2mm}
\section{Analysis}
\label{sec-ana}

\begin{figure}[ht]
\centering

\mbox{\subfigure{\includegraphics[width=5.cm]{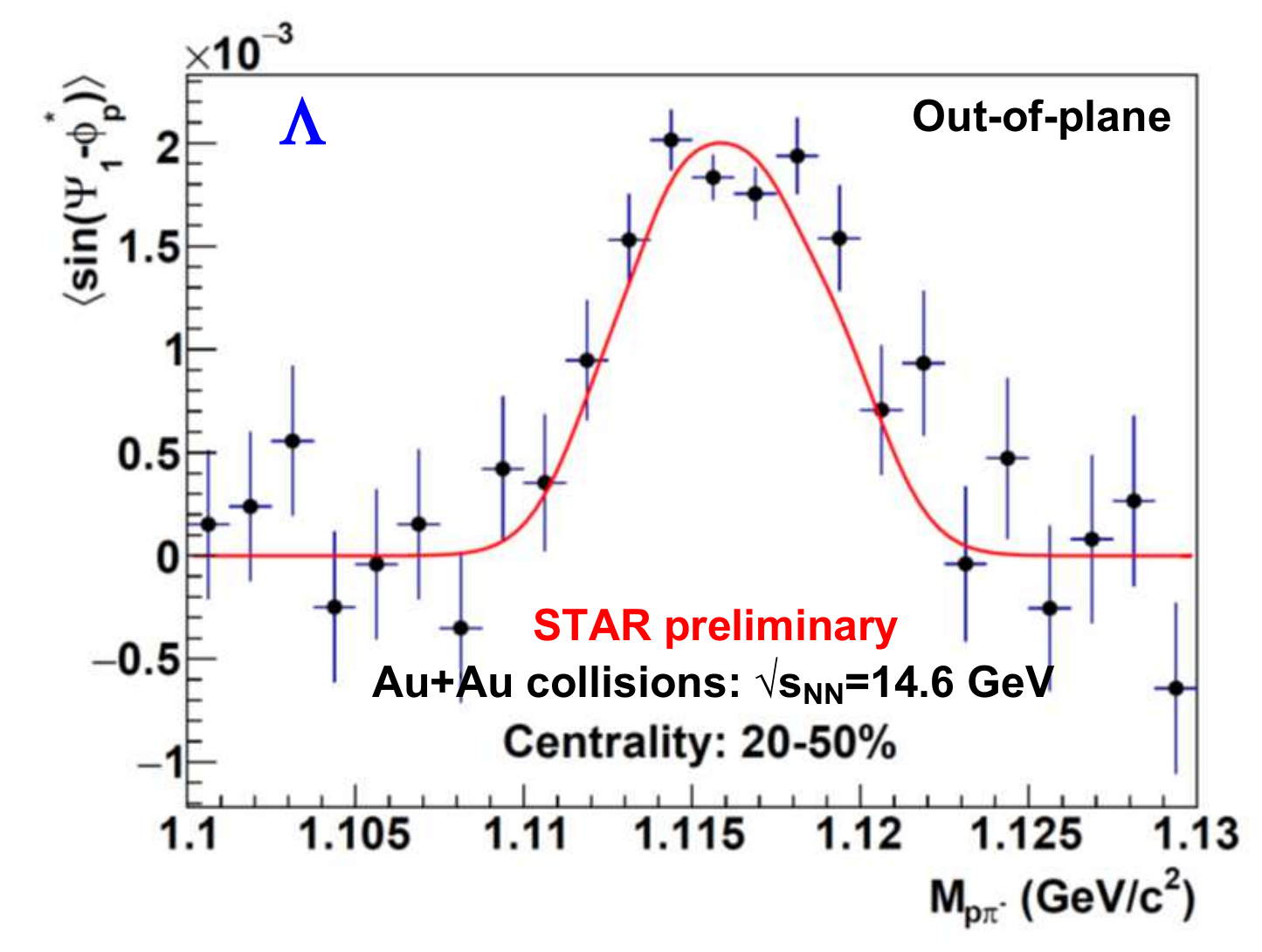}} \quad\subfigure{\includegraphics[width=5.cm]{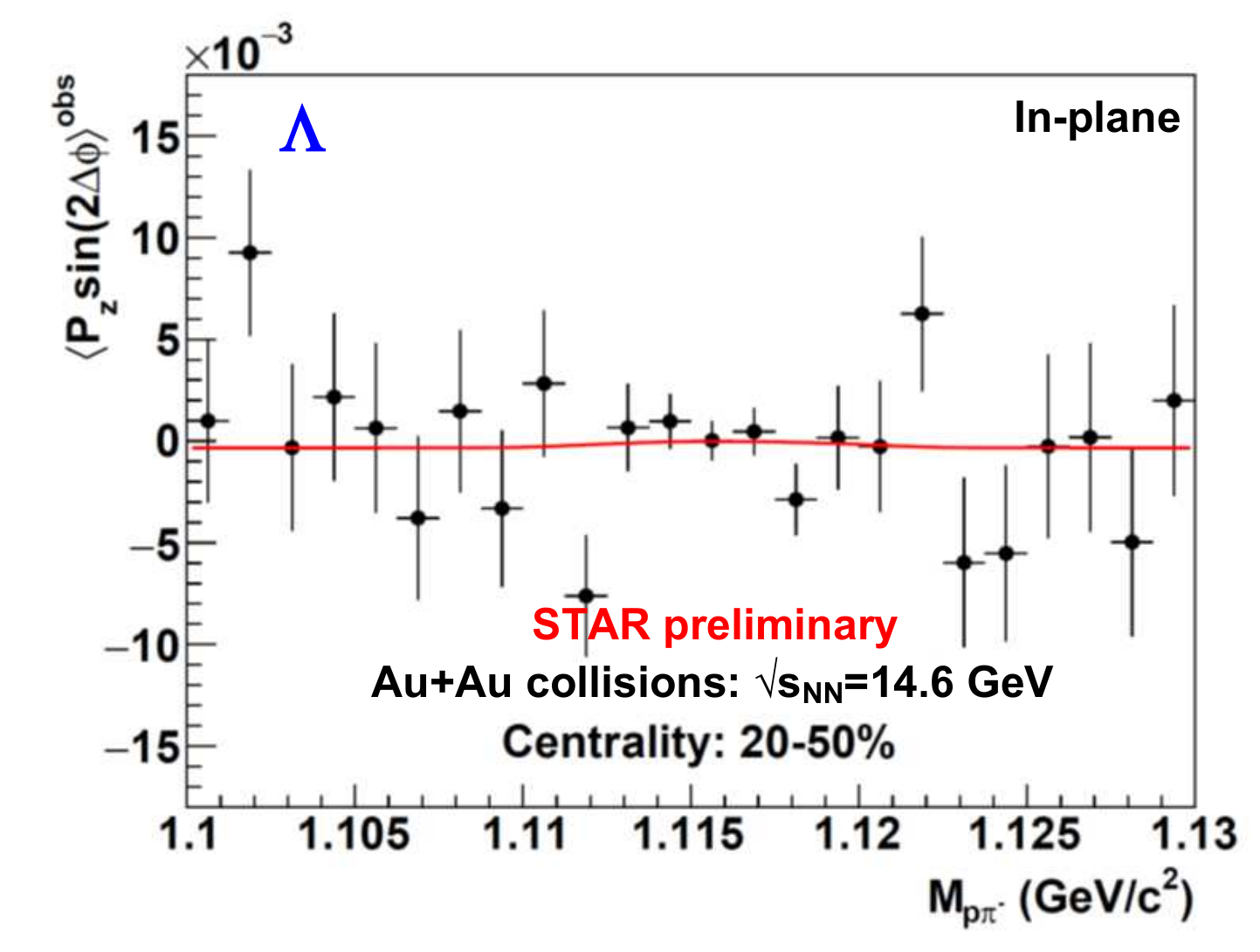}}}
\vspace*{-4mm}
\caption{$\langle\sin(\Psi_{1}-\phi_{p}^{*})\rangle$ and $\langle P_{z}\sin(2\Delta\phi)\rangle^{obs}$ as functions of invariant mass of $\Lambda$ in Au+Au collisions at $\sqrt{s_{NN}} = 14.6$ GeV. Only statistical uncertainties are shown in the plots.}
\label{fig-1}       
\end{figure}

In these proceedings, we report the global and local polarization of $\Lambda$ and $\bar{\Lambda}$ in Au+Au collisions at $\sqrt{s_{NN}}$ = 7.7, 9.2, 11.5, 14.6, 19.6, and 27 GeV, based on datasets from the second phase of the RHIC Beam Energy Scan (BES-II). The first-order event plane, $\Psi_{1}$ is reconstructed with the upgraded Event Plane Detector (EPD). The second-order event plane, $\Psi_{2}$, is determined by the Time Projection Chamber (TPC) detector. The particle identification is done with the TPC and Time-of-Flight (ToF) detector. The invariant mass spectra of $\Lambda$ and $\bar{\Lambda}$ are reconstructed with the decay channels $\Lambda \longrightarrow p+\pi^{-}$ and $\bar{\Lambda} \longrightarrow \bar{p}+\pi^{+}$. The background is evaluated by fitting the side band with the second order polynomial function. Then the ratio of signal to background ($f^{sig}$) is obtained. The distribution of $\langle \sin (\Psi_{1}-\phi_{p}^{*}) \rangle$ and $\langle P_{z}\sin(2\Delta\phi)\rangle$ (where $\Delta\phi = \phi_{\Lambda}-\Psi_{2}$,  $\langle P_{z}\rangle = \frac{\left<\cos\theta_{p}^{*}\right>}{\alpha_{\Lambda}\left<(\cos\theta_{p}^{*})^{2}\right>}$) as functions of $\Lambda$ invariant mass ($M_{p\pi^{-}}$) as shown in Fig.~\ref{fig-1}, are fitted with the expression 

$\langle P_{H} \rangle^{obs}=f^{Sig}(M_{p\pi^{-1}})\langle P_{H} \rangle^{Sig}+(1-f^{Sig}(M_{p\pi^{-1}}))\langle P_{H} \rangle^{Bg}$ ($P_{H}$ is hyperon's polarization, $\langle P_{H} \rangle^{Bg}$ is linear background in the considered invariant mass window)
to extract the signal of polarization after event plane resolution correction.

\vspace{-2mm}

\vspace*{-8mm}

\section{Results and discussion}
\label{sec-results}
\begin{figure}[ht]
\centering

\mbox{\subfigure{\includegraphics[width=5.cm]{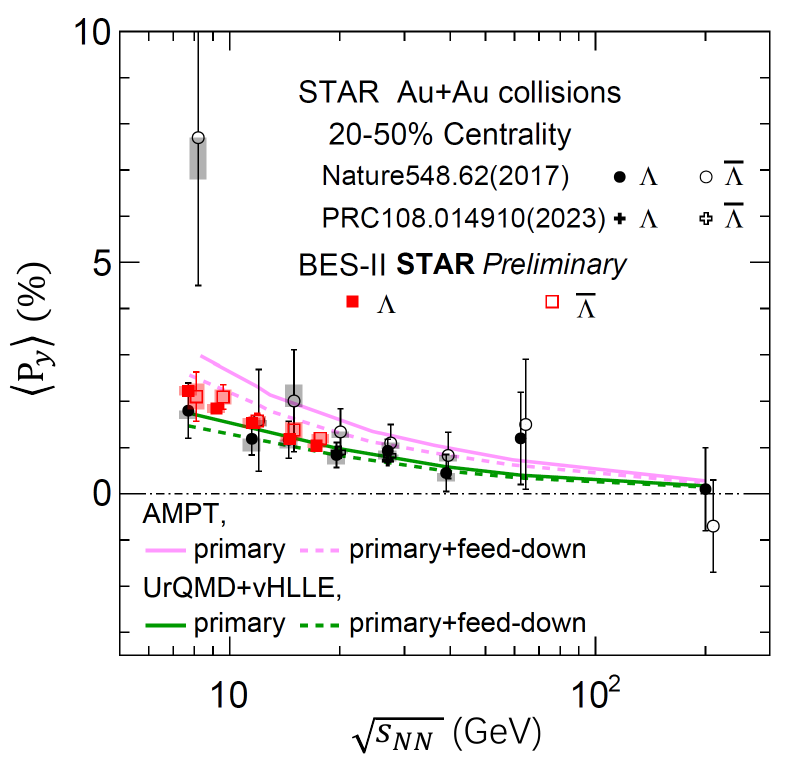}}\quad \subfigure{\includegraphics[width=5.cm]{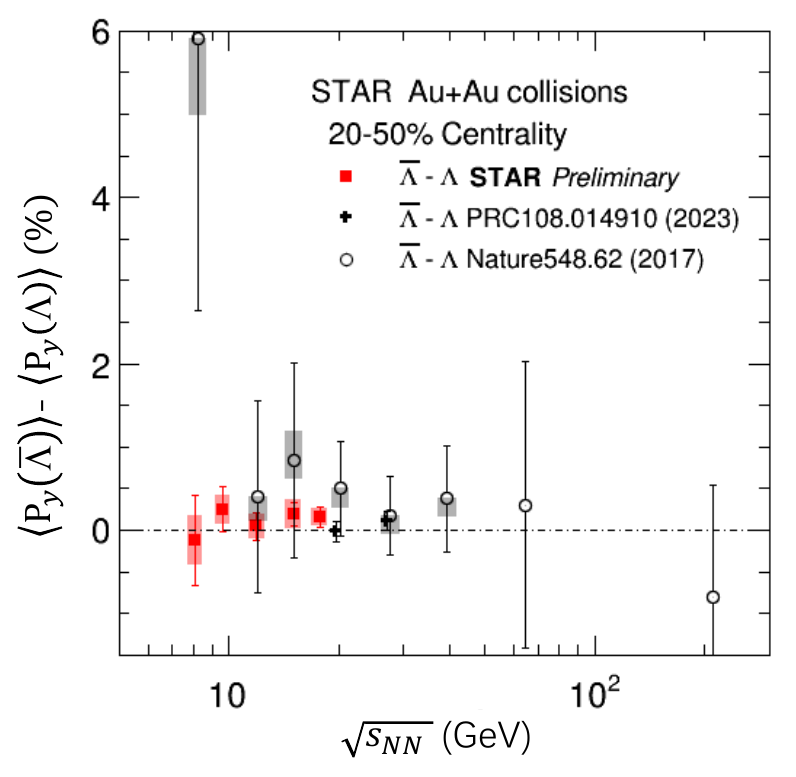}}}

\caption{(Colour online) The $\Lambda$ and $\bar{\Lambda}$ global polarization $\langle P_{y} \rangle$ (left panel) and their difference $\langle P_{y}(\bar{\Lambda}) \rangle - \langle P_{y}(\Lambda) \rangle$ (right panel) as a function of energy in Au+Au collisions. The points are data. Vertical lines and shaded boxes are statistical and systematic uncertainties.} 

\label{fig-2}      
\end{figure}

\begin{figure}[ht]
\centering

\mbox{\subfigure{\includegraphics[width=5.cm]{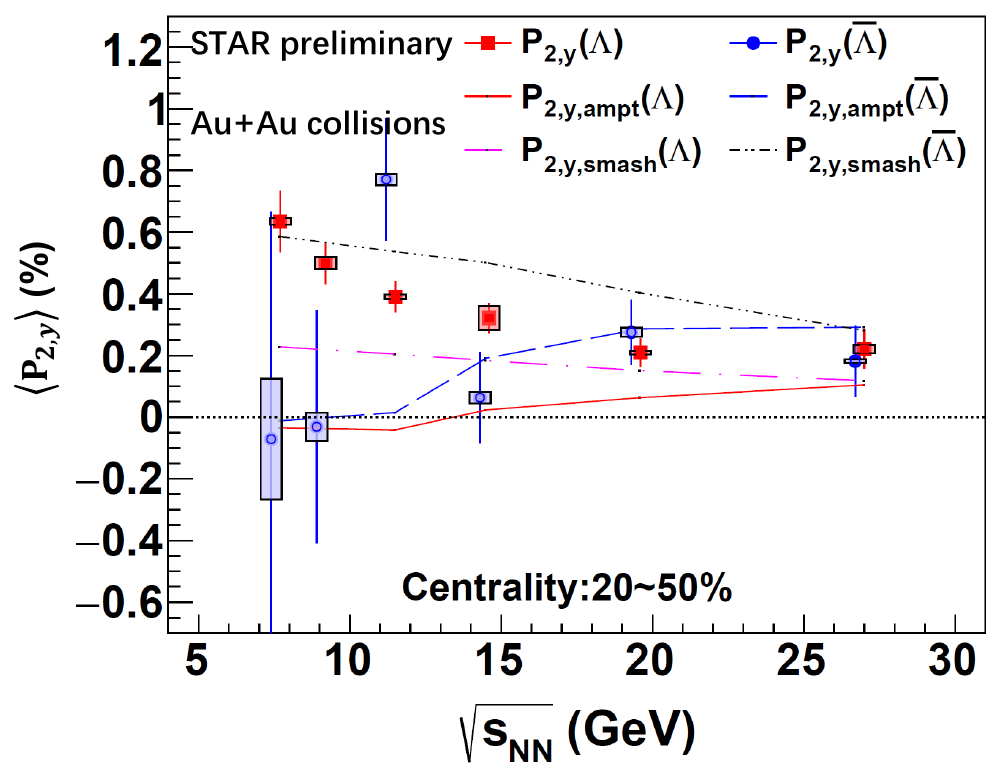}}\quad \subfigure{\includegraphics[width=5.cm]{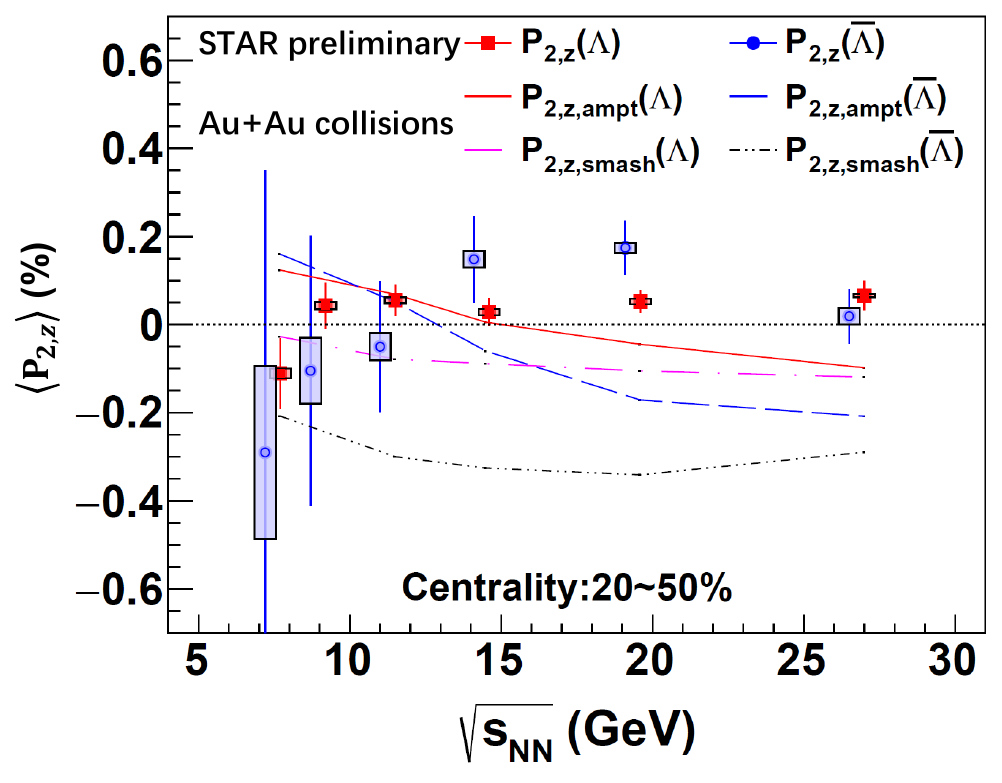}}}

\caption{(Colour online) The $\Lambda$ and $\bar{\Lambda}$ local polarization $\langle P_{2,y} \rangle$ (left pannel) and $\langle P_{2,z} \rangle$ (right panel) as a function of energy in Au+Au collisions. The red and blue points are data of $\Lambda$ and $\bar{\Lambda}$, respectively. Vertical lines and shaded boxes are statistical and systematic uncertainties.}
\label{fig-3}       
\end{figure}

\begin{figure}[ht]
\centering

\mbox{\subfigure{\includegraphics[width=5cm]{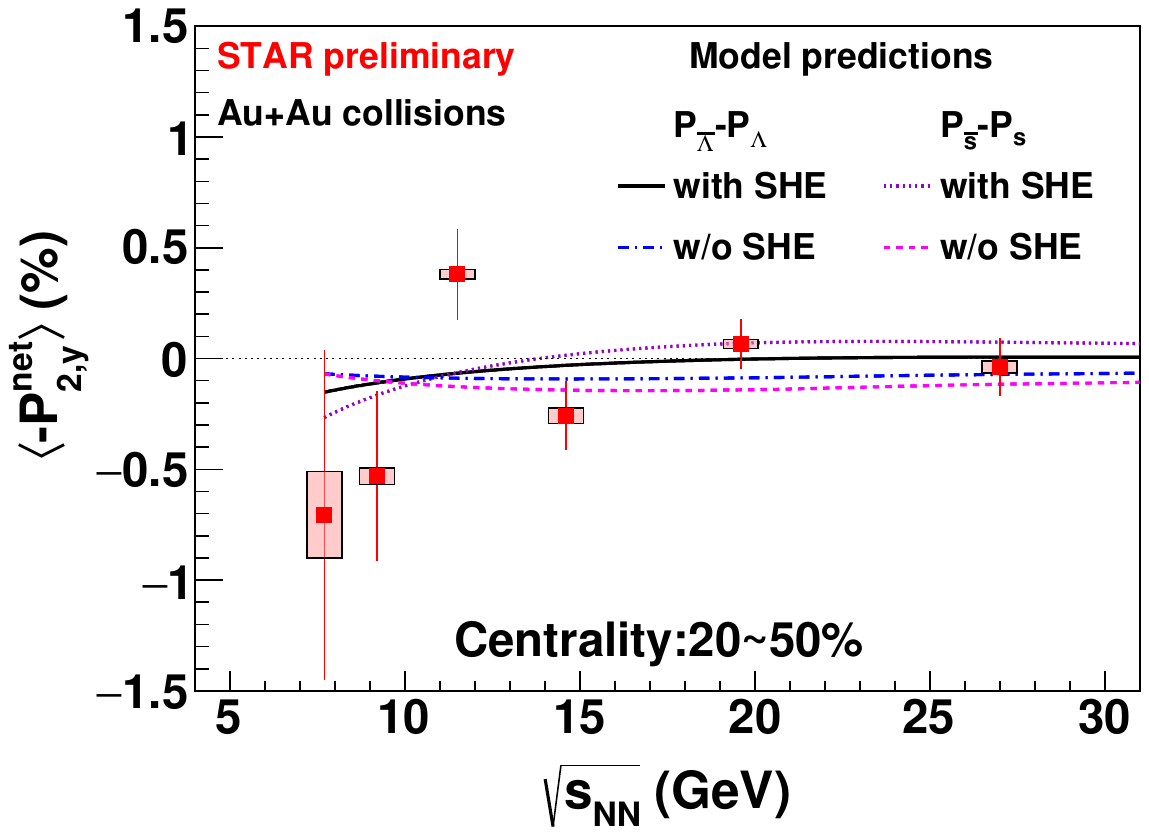}}\quad \subfigure{\includegraphics[width=5cm]{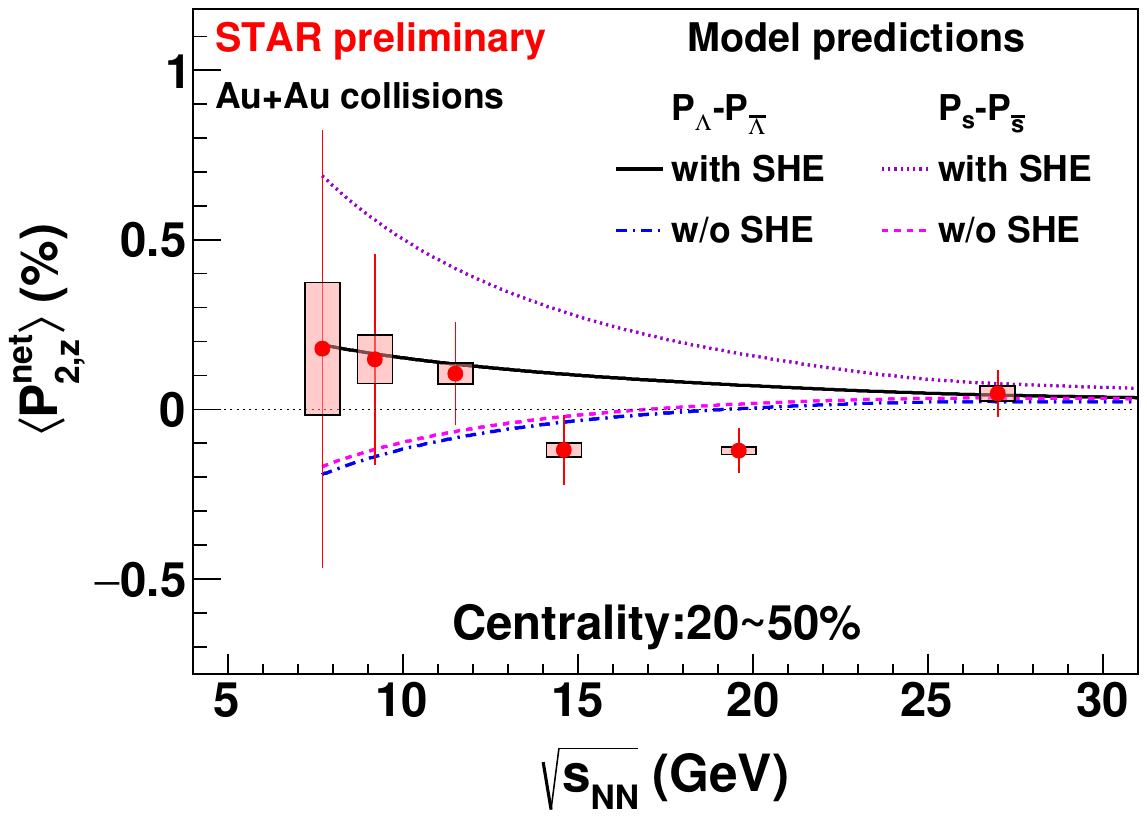}}}

\caption{(Colour online) The net local polarization $\langle -P_{2,y}^{net} \rangle$ (left panel) and $\langle P_{2,z}^{net} \rangle$ (right panel) as a function of energy in Au+Au collisions. The red solid squares are data. Vertical lines and shaded boxes are statistical and systematic uncertainties.}
\label{fig-4}      
\end{figure}

\vspace{-1mm}
Left and right panels of Fig. \ref{fig-2} present the comparison of $\Lambda$ and $\bar{\Lambda}$
global polarization and their difference, respectively. The global polarization increases as the energy decreases, and the results from BES-I and BES-II are consistent within uncertainties. The high-precision BES-II results show no significant splitting between $\Lambda$ and $\bar{\Lambda}$. Upper limits on the late-stage magnetic field at $\sqrt{s_{NN}}=$ 19.6 and 27 GeV, are estimated to be $B<9.2\times 10^{12}$ T and $B<1.4\times 10^{13}$ T, respectively \cite{B-field}. 
The left and right panels of Fig. \ref{fig-3} present the local polarization $\langle P_{2,y} \rangle$ and $\langle P_{2,z} \rangle$ for $\Lambda$ and $\bar{\Lambda}$ as a function of energy. For $\langle P_{2,y} \rangle$, the polarization of $\Lambda$ hyperons shows a clear increase with decreasing energy, while no distinct trend is observed for $\bar{\Lambda}$ within the current uncertainties. For $\langle P_{2,z} \rangle$, both $\Lambda$ and $\bar{\Lambda}$ exhibit small magnitudes and no energy dependence.
Left and right panels of Fig.~\ref{fig-4} present the net $\Lambda$ polarization, $\langle -P_{2,y}^{net} \rangle$ and $\langle P_{2,z}^{net} \rangle$, as a function of collision energy, an observable proposed to probe the SHE \footnote{In the presentation at SQM2024, the results for $\langle P_{2,y}^{net} \rangle = \langle P_{2,y}(\Lambda) \rangle - \langle P_{2,y}(\bar{\Lambda}) \rangle$ were incorrectly plotted with a factor of -1 multiplied to the data. It is corrected in these proceedings.}. No significant energy dependence has been observed for either $\langle -P_{2,y}^{net} \rangle$ or $\langle P_{2,z}^{net} \rangle$. However, $\langle P_{2,z}^{net} \rangle$ shows non-trivial trends with hints of a sign change, albeit with large uncertainties. In previous studies, global polarization could be explained by thermal vorticity, and the sign of local polarization along the beam direction could be captured by incorporating shear-induced polarization into the models \cite{fu2021,becattini2021}.

\vspace*{-5mm}

\section{Summary}
\label{sec-summ}

In a summary, we report the collision energy dependence of global and local spin polarization of $\Lambda$ and $\bar{\Lambda}$ hyperons, as well as their difference in Au+Au collisions at 7.7, 9.2, 11.5, 14.6, 19.6, and 27 GeV from the second phase of RHIC beam energy scan. With the high precision measurements, no splitting of global polarization between $\Lambda$ and $\bar{\Lambda}$ has been observed. The upper limit of late stage magnetic field at 19.6 and 27 GeV have been estimated. The local polarization of $\Lambda$ and $\bar{\Lambda}$ ($\langle P_{2,y} \rangle$, $\langle P_{2,z} \rangle$) are measured for the first time at RHIC BES energies. A monotonic energy dependence of $\Lambda$ local polarization along out-of-plane direction has been observed. The net local polarization perpendicular to the reaction plane and along the beam directions have been measured. The newly proposed mechanism of the baryonic spin Hall effect has been probed using the local polarization difference between $\Lambda$ and $\bar{\Lambda}$, and no indication of SHE is observed within the current precision.
However, it brings challenges to the hydrodynamic models to describe $\langle P_{2,y} \rangle$ and $\langle P_{2,z} \rangle$ simultaneously with or without SHE.
More theoretical input is needed to understand $\Lambda$ polarization at high baryon densities.

\section*{Acknowledgments}
This work is supported in part by the National Key Research and Development Program of China under Contract No. 2022YFA1604900.

\end{document}